\documentclass[twocolumn,showpacs,amsmath,amssymb,prb]{revtex4}

\usepackage{amsmath}
\usepackage{amssymb}
\usepackage{graphicx}
\usepackage{dcolumn}
\usepackage{bm}
\usepackage{color}
\def\be{\begin{eqnarray}}
\def\ee{\end{eqnarray}}

\newcommand{\n}{\textcolor{blue}}

\begin{document}
\title{Anomalous resistance overshoot in the integer quantum Hall effect}

\author{E. M. Kendirlik$^1$}
\author{S. Sirt$^1$}
\author{S. B. Kalkan$^1$}
\author{W. Dietsche$^2$}
\author{W. Wegscheider$^2$}
\author{S. Ludwig$^3$}
\author{A. Siddiki$^1$}
\address{$^1$Department of Physics, Istanbul University,
Istanbul, 34134 Turkey}
\address{$^2$Laboratory for Solid State Physics, ETH Z\"urich, CH-8093 Z\"urich, Switzerland}
\address{$^3$Center for NanoScience and Fakult\"at f\"ur Physik,
Ludwig--Maximilians--Universit\"at, Geschwister--Scholl--Platz 1,
D--80539 M\"unchen, Germany}
\begin{abstract}
In this work we report experiments on defined by shallow etching narrow Hall bars. The magneto-transport properties of intermediate mobility two-dimensional electron systems are investigated and analyzed within the screening theory of the integer quantized Hall effect. We observe a non-monotonic increase of Hall resistance at the low magnetic field ends of the quantized plateaus, known as the overshoot effect. Unexpectedly, for Hall bars that are defined by shallow chemical etching the overshoot effect becomes more pronounced at elevated temperatures. We observe the overshoot effect at odd and even integer plateaus, which favour a spin independent explanation, in contrast to discussion in the literature. In a second set of the experiments, we investigate the overshoot effect in gate defined Hall bar and explicitly show that the amplitude of the overshoot effect can be directly controlled by gate voltages. We offer a comprehensive explanation based on scattering between evanescent incompressible channels.
\end{abstract}
\date{\today}
\maketitle
\section{INTRODUCTION}
\label{sec:intro}
The overwhelming interest to utilize quantum mechanics in applied technologies finds one of its first manifestations in the integer quantized Hall effect (IQHE).~\cite{Klitzing:1980ti} The magnetic field dependence of the transport coefficients of a two dimensional electron system (2DES) provides a possibility to standardize resistance in units of the von Klitzing constant $h/e^2$, where $h$ is the Planck constant and $e$ is the elementary charge. However, an unexpected non-monotonic magnetic field dependence of the Hall resistance at the low-field-end of the quantized plateaus, known as the overshoot effect, remains a puzzle despite of both theoretical and experimental efforts in various material systems including GaAs/AlGaAs heterostructures~\cite{Zheng:1986bs,Alphenaar:1990bs,McEuen:1990bs,Richter:1992ck,Komiyama:1993ck} as well as  Si/SiGe~\cite{Coleridge:1996bs,Dunford:2000bs,Shlimak:2000bs,Shlimak:2004bs,Shlimak:2005bs,Sailer:2010bs,Payette:2012bs} and Si metal oxide semiconductor field effect transistors.~\cite{Shlimak:2006bs} The utilization of the quantized Hall effect as a resistance standard is hindered by such anomalies, especially because their physical mechanism is not well understood. The overshoot effect is observed in these material systems at various filling factors $\nu$, defined by the number of occupied quantized (spin resolved) Landau levels (LL) below the Fermi energy. The effect has already been observed in the 1980's,  where its  physical mechanism was attributed to non-ideal contacts,~\cite{Zheng:1986bs,Alphenaar:1990bs,McEuen:1990bs} but without providing clear evidence for this hypothesis. Later, the overshoot effect was attributed to the decoupling of the spin-split states within the same LL at odd filling factors by Richter and Wheeler,~\cite{Richter:1992ck} or, alternatively by the scattering between edge states together with spin-orbit interaction by Komiyama and Nii.~\cite{Komiyama:1993ck} Recently, the overshoot effect has been investigated in Si/SiGe heterostructures as a function of current and temperature.~\cite{Sailer:2010bs} These experimental results have been elegantly explained within the screening theory of the integer quantized Hall effect, which explicitly takes  into account the direct Coulomb interaction between charge carriers. In this approach the overshoot effect is described using co-existing (current carrying) evanescent incompressible strips,~\cite{Siddiki:2010bs1} while earlier explanations used 1D Landauer-Buttiker edge channels.~\cite{Alphenaar:1990bs} Under certain conditions, namely when an incompressible strip is narrower than the Fermi wavelength, but wider than the magnetic length, the carriers can scatter between adjacent evanescent incompressible regions causing an increase in the Hall resistance. This situation resembles a leaky incompressible strip in the thermodynamical sense, which then carries a dissipative current. Such an incompressible strip will be called \emph{evanescent} throughout the paper. A detailed theoretical explanation of the overshoot effect within the screening theory is provided in Ref. 18 taking into account finite size and temperature effects. Ref. 18 specifically predicts that the overshoot effect can be manipulated by changing the electrostatic edge profile of the electron gas, for example by utilizing side gates.

Here, we present experiments on narrow Hall bars $ (\leq 10 \mu$m) which are defined by either shallow chemical etching, or metallic gates employing the field effect in GaAs/AlGaAs heterostructures. Metallic gates provide the possibility of controlling smooth edge potential profiles,~\cite{Horas:2008bs,Siddiki:2009bs,Siddiki:2010bs} perfect for testing the predictions outlined above: The overshoot effect is predicted to vanish if the co-existence of evanescent (leaky) incompressible strips is destroyed by a steep potential at the edge, or is enhanced by smooth potentials in the opposite limit.

The paper is organized as follows: We start with a theoretical introduction summarizing the concepts of the screening theory. Then, we present the results of Hall resistance measurements on 4 and 10 $\mu$m wide shallow etch defined Hall bars (Samples IA, and IB, respectively). Another set of experiments is based on a gate defined 3 $\mu$m wide Hall bar (Sample IIA). We will compare our measurements with the predictions of the screening theory and discuss the implications regarding the overshoot effect.

\section{THEORETICAL BACKGROUND}

At sufficiently low temperatures and high magnetic fields, the direct Coulomb interaction separates a 2DES into compressible and incompressible regions of finite lateral size with very different screening properties. Their theoretically predicted existence has been investigated in various experiments including electrostatic transparency and dynamical scanning capacitance measurements.~\cite{Ahlswede:2002bs,Suddards:2012bs} The theoretical prediction of the existence of compressible and incompressible strips dates back to 1990, propounded by Chang.~\cite{Chang:1990ck} This work was followed by a pioneering paper of Chklovskii and co-workers who calculated the widths and spatial distributions of these strips analytically.~\cite{Chklovskii:1992ti} The formation of the strips can be traced back to a stepwise electron density distribution. In the commonly employed single particle picture, the LL are bend up in energy at the edges of the 2DES and are filled up to the Fermi energy (at T=0). At $\nu\geq1$ starting from the edges the lowest LL is completely occupied hence contributing to the carrier density with a constant value (at fixed $B$). Moving from the edges of the Hall bar towards its center the carrier density changes stepwise whenever a LL crosses the Fermi energy. The situation is further modified when taking into account the electron-electron interaction. A stable solution is found by minimizing the free energy while considering the Coulomb interaction between the carriers. The result are regions of varying carrier density profile  (the compressible strips), where the total potential is flat, and regions of constant carrier density profile (the incompressible strips), where the total potential varies. The widths of the $k^{\rm th}$ incompressible strip (with local filling factor $k$) can be evaluated up to a reasonable approximation by an analytic formula~\cite{Siddiki:2010bs}
\be a_k=\Large(\frac{2\kappa \Delta E_k}{\pi^2e^2dn(x)/dx|_{x_k}}\Large)^{1/2},\label{widths} \ee
where $\kappa$ is the dielectric constant ($\sim 12.4$, for GaAs), and $n(x)$ is the electron density at $B=0$ as a function of lateral coordinate $x$. Here, the density gradient is evaluated at the center of the $k^{th}$ incompressible strip, $x_k$. The single particle gap $\Delta E_k$ is the extra energy (in addition to the chemical potential at $B=0$) needed to load another electron into the system. It consists of the cyclotron energy $\hbar\omega_c=\hbar eB/m^*$ and the Zeeman energy $g^*\mu_BB$, where $\mu_B$ is the Bohr magneton and $g^*$ is the effective Land\'e $g$ factor. For odd (even) filling factor the energy gap is $\Delta E_{odd}=g^*\mu_BB$  ($\Delta E_{even}=\hbar\omega_c-g^*\mu_BB$). The local carrier density distribution at zero magnetic field can be obtained within self-consistent numerical calculations~\cite{Salman:2013ti}:
\be n(x)=n_0(1-e^{-(x-l_d)/t}), \label{density}\ee
where $l_d$ is the depletion length and $n_0$ is the bulk electron density far away from the edges. The parameter $t$ defines the distance from the edge at which the electron density reaches $n_0$, in units of the effective Bohr radius $a_B^*$.
Substituting Eq. \ref{density} into Eq.~\ref{widths}, one obtains the incompressible strip width
\be a_k=\sqrt{\frac{4a_B^*\alpha_k}{\pi \nu_0}}\times\sqrt{\frac{t}{e^{-(x_k-l_d)/t}}}, \label{eq:ak_width}\ee
where $\alpha_k=\Delta E_k/\hbar\omega_c$ is the gap parameter. The bulk filling factor defined at the center of the Hall bar is $\nu_0=\pi\ell^2n_0$ with the magnetic length $\ell=\sqrt{\hbar/eB}$.

In the above calculation we assumed that the Thomas-Fermi approximation (TFA) is valid, i.e. the electrostatic potential varies smoothly on the scale of $\ell$. However, once the strip widths become comparable with the magnetic length $a_k\lesssim \ell$ the TFA is prone to fail.~\cite{Chklovskii:1992ti} At this point scattering across the strip becomes more probable.~\cite{Suzuki:1993ck,Siddiki:2004ck} Furthermore, the electron density and compressibility are thermodynamic quantities which are only properly defined for length scales larger than the mean electron distance. In the low temperature limit this is the Fermi wavelength $\lambda_F$ (hence for $a_k\lesssim \lambda_F$ compressibility is not a well defined quantity). The above discussion yields a lower bound for $B$ the width at which an incompressible strip can exist. Each incompressible strip becomes narrower with decreasing $B$ (see Eq.~\ref{eq:ak_width}) and for $a_k\lesssim \lambda_F$ it eventually becomes thermodynamically permeable (or leaky). This process results in the transition regions between subsequent Hall plateaus. For $\ell \lesssim a_k\lesssim \lambda_F$ we call an evanescent incompressible strip evanescent.~\cite{Sailer:2010bs} The scattering model predicts the overshoot effect to occur if at least two evanescent incompressible strips with consecutive filling factors co-exist: $\ell<(a_k,a_{k+1})\lesssim \lambda_F$.~\cite{Siddiki:2010bs} To be explicit, if (at least) two incompressible strips with different filling factors are narrower than $\lambda_F$, and wider than $\ell$ (suppressing scattering across them) both of the channels contribute to the imposed current, resulting in an increase of the Hall resistance. For a hand waving (and simplified) example we assume a co-existence of two evanescent strips with $\nu=2$ and $\nu=3$, which share the imposed current equally, i.e. $I_2=I_3=I/2$. Assuming no additional dissipation the resulting resistance
\be {R_H}= \frac{h}{e^2}\times\frac{1}{2}(\frac{1}{2}+\frac{1}{3}) \\
=\frac{5}{12}\frac{h}{e^2},\ee
which is larger than the quantized value of $\frac{1}{3}e^2$, while this should not be taken as a quantitative prediction it sketches the general situation which we will observe in the following sections.

\begin{figure*}[ht!]
{\centering
\includegraphics[width=1.\linewidth]{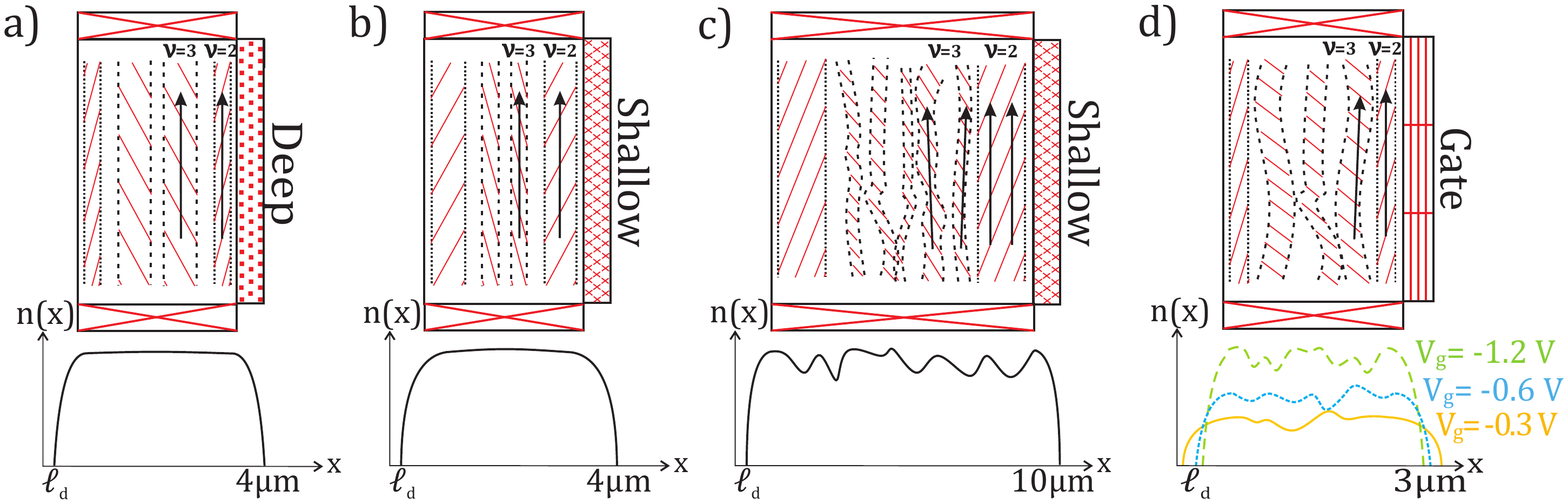}
\vspace{-7cm}
\caption{\label{fig1}(Color online) A schematic presentation of the electron density as a function of the lateral coordinate together with the evanescent incompressible strips indicated by dotted ($\nu=2$) and dashed ($\nu=3$) vertical lines carrying dissipative current (depicted by arrows). the two left sketches show 4 $\mu$m wide Hall bars defined by (a) deep and (b) shallow etching. The density oscillations results from long-range potential fluctuations due to remote donors. (c) 10 $\mu$m wide Hall bar comprising several long-range disorder induced density oscillations, resulting in several bulk incompressible strips dominating the transport, hence overshoot. (d) Gate defined narrow Hall bar, the lower panel sketches electron densities for 3 different gate voltages, where the depletion length changes and the maximum of the electron density increases at higher gate voltages, however, the average density remains almost the same. Note that drawing are not to scale.}}
\end{figure*}

One can readily see from Eq.~\ref{widths} that, by manipulating the edge potential (or equivalently the edge density) profile it is possible to obtain wide (large $t$) or narrow (small $t$) incompressible strips. Interestingly, depending on the energy gap and steepness it is also possible to obtain conditions such as, $\ell<a_k < a_{k+1}\lesssim \lambda_F$ or $\ell< a_{k+1}< a_k\lesssim \lambda_F$. For instance one can obtain a situation $a_1>a_2$ if $\Delta E_1>\Delta E_2$ with an exchange enhanced $g^*$ factor, for $t\gtrsim 6 a_B^*$ defining a smooth edge (c.f Eq.~\ref{eq:ak_width}).

Experimentally Hall bars can be defined by etching or by depositing gates on the surface. In the case of etching the crystal is usually removed beyond the 2DES plane, for this so called deep etching, the confinement potential at the edges becomes steep due to surface charges inside the etched tranches.~\cite{Halperin:1982ck} This situation corresponds to the small $t$ limit, which is most common for Hall bars, Fig.~\ref{fig1}a. In the limit of shallow etching, where the crystal is only removed above the 2DES plane, the confinement is relatively flat, as depicted in Fig.~\ref{fig1}b. According to the discussion above a smoother confinement potential as that arising from shallow etching results in a higher probability of overshoot effects. Gated samples have the advantage that the edge profile can be adjusted via the gate voltages between steep and flat edges on one and the same Hall bar. In the next section, we will discuss magneto-transport experiments first on shallow etched Hall bars, and then on a gate defined Hall bar and compare the results with our model.

\section{EXPERIMENTAL RESULTS}
\label{sec:exp}

We performed standard magneto-transport measurements on narrow Hall bars defined in GaAs/AlGaAs heterostructures. The first set of Hall bars are defined by shallow chemical etching (samples IA and IB), whereas an additional sample is defined by metallic gates (sample IIA). Samples IA and IB differ in their widths, 4 $\mu$m and 10 $\mu$m respectively (IIA has 3$\mu$m width). The used wafer contains a 2DES approximately 100 nm below the surface while the etching depth was about 80 nm. The nominal mobility of the wafer is 380000 cm$^{2}$/Vs at an electron density of $2.45\times 10^{11}$ cm$^{-2}$. The gate defined sample is produced on a heterostructure with the 2DES 110 nm below the surface, with an electron density of $2.8\times 10^{11}$ cm$^{-2}$ and a nominal mobility of $1.4\times 10^{6}$ cm$^{2}$/Vs. More details of the gate defined Hall bars can be found in Refs. [16-18].

\subsection{Etched samples below 1 K}

In the first set of experiments (on samples IA and IB) we measured the Hall voltage using standard lock-in technique at a frequency of 8.5 Hz, as a function of magnetic field at temperatures below 1K. We start with the narrower sample IA. Magnetotransport properties in the quantized Hall regime are mainly determined by the edges of the sample including the widths of the incompressible strips. Fig\n{.} 2 depicts the Hall resistance of the 4 $\mu$m wide sample IA. In addition to the integer quantized Hall plateaus between $2\leqslant\nu\leqslant6$, the overshoot effect is clearly present at the low-field end of the $\nu=3$ plateau. The amplitude of the overshoot effect increases at elevated temperatures, but decreases with increasing excitation current at a fixed temperature (of 750 mK, inset). Interestingly, these experimental findings are in clear contrast to literature expectations,\textbf{~\cite{Sailer:2010bs}} but in agreement with the screening theory as elucidated below. Note that we ruled out sample dependencies and the effect of a trivial contact resistance by careful comparison of measurements before and after illumination, by performing several cool downs and by testing various contact configurations.

\begin{figure}[t]
{\centering
\includegraphics[width=1.\linewidth]{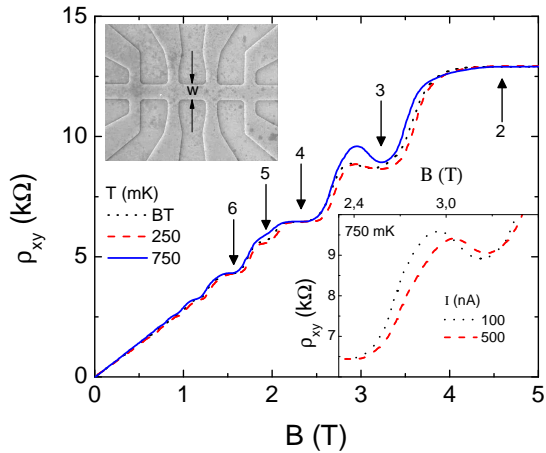}
\vspace{0cm}
\caption{\label{fig2}(Color online) The Hall resistivity measured at a 4 $\mu$m wide shallow etch defined Hall bar as a function of magnetic field measured at three different temperatures, base temperature BT (dotted line), 250 mK (broken line) and 750 mK (solid line). The excitation current amplitude is fixed to 100 nA. A well developed overshoot effect is observed at $\nu=3$ plateau, which becomes more pronounced at elevated temperatures. The inset depicts current amplitude dependency at 750 mK. The topographic image of the sample is shown at the upper inset, where $W$ denotes the width.}}
\end{figure}

The observed temperature and current dependence of the overshoot effect can be explained within the screening theory as follows: The mere existence of the overshoot effect is caused by the co-existence of two evanescent incompressible strips in our case at local filling factors $\nu=2$ and $\nu=3$. According to Eq. 1 the width of each incompressible strip is proportional to $\sqrt{\Delta E_k}$, which alternates between even and odd filling factors, and also inversely proportional to the square root of the lateral carrier density gradient $\frac{dn(x)}{dx}$. In a sample with weak disorder $\frac{dn(x)}{dx}$ is always smaller for the inner one of two co-existing incompressible strips (see Eq.~\ref{density}), which also has the higher filling factor. The ratio of the widths of two co-existing incompressible strips can be calculated from Eq.~\ref{eq:ak_width} as
\be \frac{a_n}{a_m}=\sqrt{{\frac{\alpha_n}{\alpha_m}}\times{e^{-\frac{x_m-x_n}{t}}}}\quad. \label{an:am_width}\ee
where we take $m=n+1$ for co-existing strips. For odd (even) $n$ we find $\sqrt{\frac{\alpha_n}{\alpha_m}}\simeq8.1$. This leads to the interesting possibility to change the ratio $\frac{a_n}{a_m}$ via the edge parameters $t$ which is larger (smaller) for edges defined by deep (shallow) etching and can be adjusted for a gate defined edge. For our specific case (Fig\n{.} 2) we argue that the overshoot is caused by a co-existence of evanescent incompressible strips of filling factors $\nu=2$ and $\nu=3$, hence $n=2$ and $m=3$. From Eq.~\ref{an:am_width} we find $a_2\leq a_3$ for $t\geq \frac{x_3-x_2}{4}$.

In the common case of deeply etched Hall bars we expect $t< \frac{x_3-x_2}{4}$ and the bulk strip ($\nu=3$) to be wider, see Fig\n{.} 1b. As a consequence along the $\nu=3$ plateau electron transport would always be dominated by the $\nu=3$ strip and no overshoot was expected. However, the data in Fig\n{.} 2 have been measured in a shallow etched sample and the existence of the overshoot points to $t> \frac{x_3-x_2}{4}$. As a consequence, at the low field end of the $\nu=3$ plateau the bulk strip is narrower than the edge strip ($\nu=2$) and the transport properties are influenced by the $\nu=2$ incompressible strip. This can lead to an increase of the resistance beyond the $\nu=3$ and up to the $\nu=2$ plateau value. In more detail we expect an overshoot for $a_3 <l$ and $l<a_2<\lambda_F$.

According to the above arguments overshoots can be expected only (never) at the low field end of odd (even) filling factor plateaus. The reason is the alternating gap size $\Delta E_{even}\gg \Delta E_{odd}$. In Fig\n{.} 2 we observed that the overshoot increases with growing temperature. This is in accordance with our model assuming that the narrower bulk strip is stronger affected by temperature (and becomes easier compressible) compared to the wider edge strip. Hence, at higher temperature the influence of the $\nu=2$ edge strip increases towards the $\nu=2$ plateau value up to an even higher temperature where the edge strip also becomes completely compressible. When this happens the overshoot resistance should decrease again (towards the classical Hall resistance). In the present experiment we could not observe the high temperature limit due to a technical restriction, but measured instead the current dependence at the highest temperature of 750 mK.  Since in the overshoot regime the current is dissipative, due to Joule heating the imposed current warms up the current carrying strip. Due to the fact that the $\nu=2$ evanescent strip has a higher resistance, it warms up more and breaks easier than the $\nu=3$ strip and the overshoot decreases.

To summarize, on a 4 $\mu$m wide shallow etched sample we observed the overshoot effect at the low field end of the $\nu=3$ plateau. Our model indicates that the edge $\nu=2$ evanescent incompressible strip is wider than the bulk $\nu=3$ strip in contrast to literature prediction. The effect is more pronounced at elevated temperatures. This behavior can be explained by  the fact that the inner thinner $\nu=3$ strip vanishes before the outer wider $\nu=2$ strip because of $\Delta E_{even}> \Delta E_{odd}$. In the extreme case where the $\nu=3$ strip already brakes down and the $\nu=2$ strip still exists, the Hall resistance can even approach to $\frac{1}{2}\frac{h}{e^2}$ (with dissipative corrections). At higher imposed currents due to dissipation proportional to the local resistance, the $\nu=2$ evanescent incompressible strip vanishes rapidly yielding a decrease of the overshoot, driven by the complete breakdown of the quantized Hall effect.

\begin{figure}[t]
{\centering
\includegraphics[width=1.\linewidth]{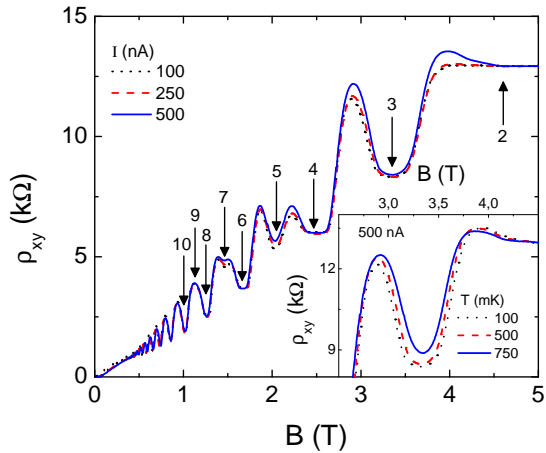}
\caption{\label{fig3}(Color online) The Hall resistivity measured at a 10 $\mu$m width, shallow etch defined Hall bar as a function of the magnetic field measured at three different excitation amplitudes at 100 mK. The overshoot effect is observed at both even and odd integer plateaus, due to different scattering mechanisms, as discussed in the text. The inset shows the temperature dependency of the $\nu=3$ overshoot considering different temperatures.}}
\end{figure}

Next we study a relatively wide Hall bar (10 $\mu$m) defined on the same wafer again by shallow etching. Fig.~\ref{fig3}, depicts the Hall resistance as a function of the $B$ field. Here, the overshoot effect is very strong at the low-field end of the $\nu=3$ plateau, but also appears on the other plateau ($\nu\geq2$). The overshoot for odd $\nu$ can be explained within the model discussed above. However, it also appears at even $\nu$ which requires further discussion: We interpret the occurrence of the overshoots at even $\nu$ in terms of disorder in the bulk: local potential fluctuations (in space, not in time) add to the potential drop induced by the edges. If two incompressible strips coexist the influence of disorder is stronger at the inner (bulk) incompressible strip where the edge profile is already less steep compared to the outer (edge) incompressible strip. In this situation disorder can result in a strong enhancement of the gradient $\frac{dn(x)}{dx}$ and, consequently, a thinner bulk strip.

We start our discussion with the $\nu=2$ plateau, for which we observe an overshoot for the largest imposed current ($I$=500 nA, solid line), however, a quantized Hall resistance for lower currents (broken and dotted lines). In this situation the bulk incompressible strip with $\nu=2$ is well developed and stays stable at low currents. However, it becomes evanescent due to increased potential drop across the strips at larger currents. In addition due to its exchange enhanced Zeeman gap, the outer incompressible strip with $\nu=1$ satisfies the condition $\lambda_F>a_1>\ell$, hence is evanescent. Therefore, we observe an overshoot only at large currents, where both incompressible strips become evanescent.

We now discuss the case of even $\nu>2$: once the sample is sufficiently wide to accommodate more than a couple of long-range potential fluctuations, indicated by the density modulation in $n(x)$ Fig~\ref{fig2}c, the bulk dominates the scattering mechanism yielding more than one \emph{bulk} evanescent incompressible strip. Hence, for a shallow etched sample, an even \emph{edge} integer evanescent incompressible strip can co-exist with the odd evanescent incompressible \emph{bulk} strips. We expect that the disorder is more effective in the large sample, since it offers more long-range potential fluctuations leading to a bulk dominated transport.~\cite{Siddiki:2010bs1} We also determined the exact bulk filling factors from the SdH oscillations and checked for the coincidence of the maximum of the overshoot effect and the bulk filling factor. The mechanism is the same, however, the effect is more immune to heating effects due to dissipation, since there exists many bulk strips which share the total current.

\begin{figure}[t]
{\centering
\includegraphics[width=1.\linewidth]{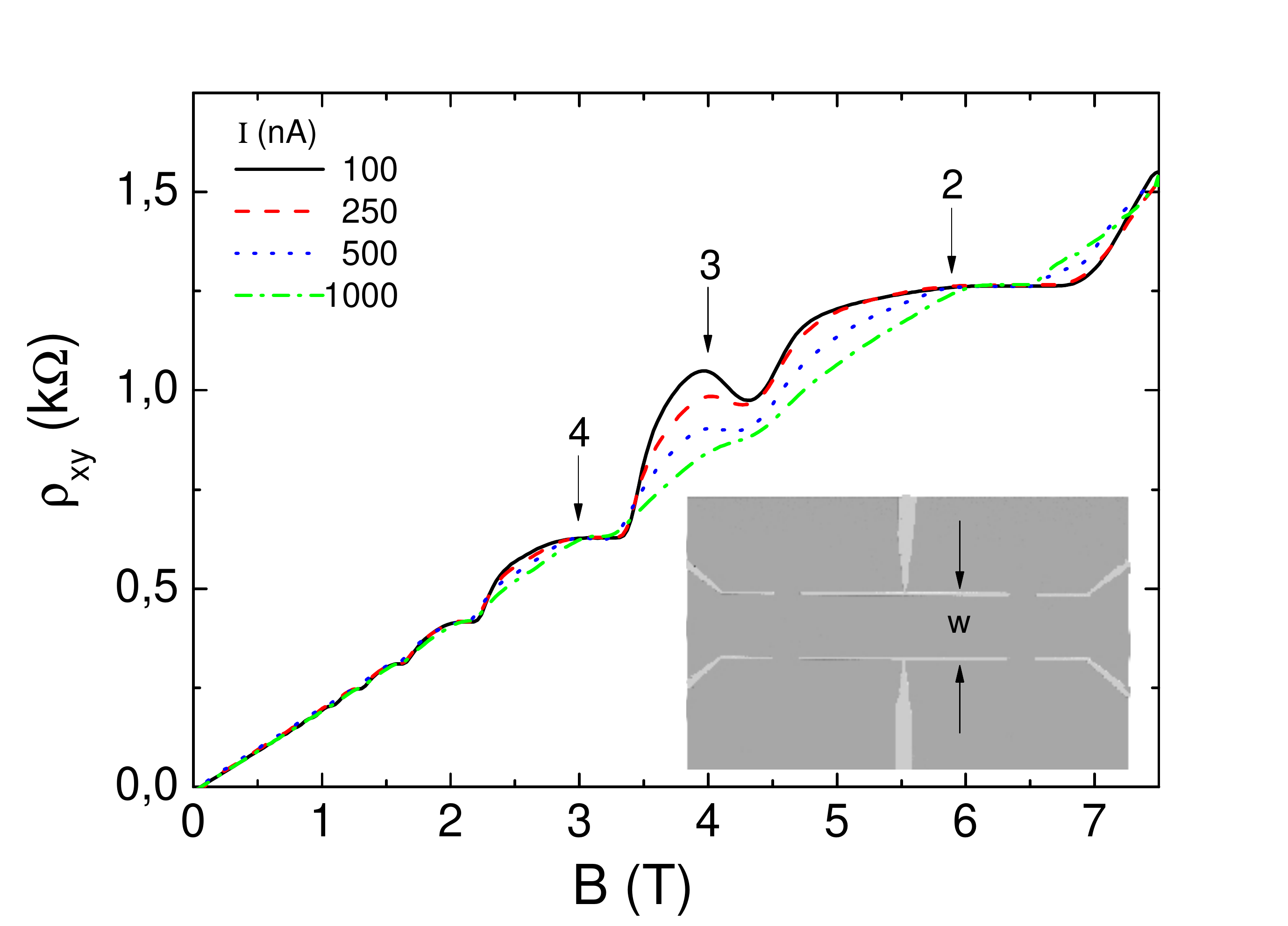}
\vspace{0mm}
\caption{\label{fig4}(Color online) The Hall resistance measured at a gate defined 3 $\mu$m width Hall bar at 1.7 K, while imposing different excitation currents. The $\nu=3$ overshoot fades with increasing the current amplitude, where edges are supposed to be smooth due to the small gate bias voltage of -0.3 V.}}
\end{figure}

\subsection{Gated samples above 1K}
\begin{figure}[t]
{\centering
\includegraphics[width=1.\linewidth]{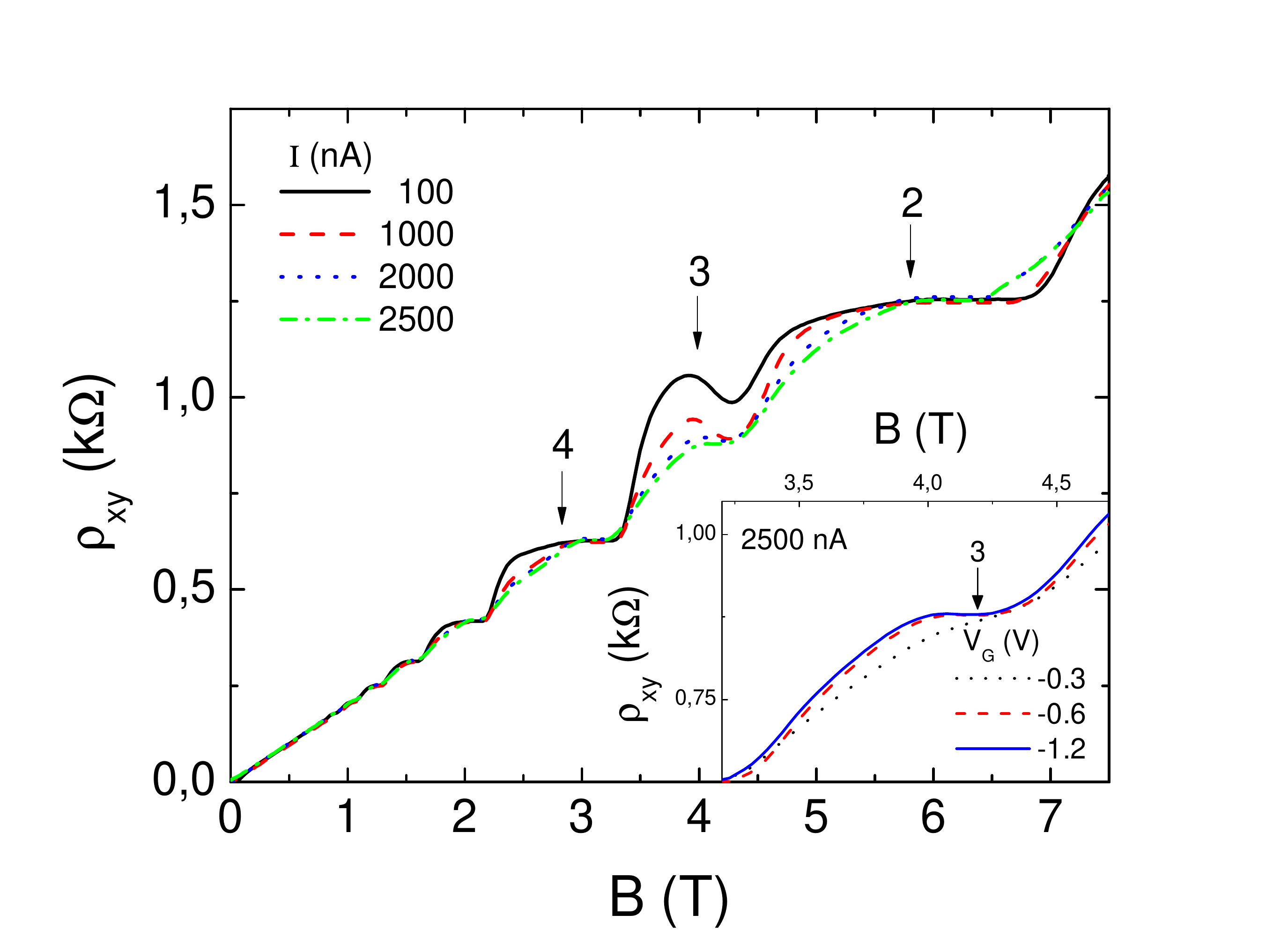}
\vspace{0mm}
\caption{\label{fig5}(Color online) Same as Fig.~\ref{fig4}, where the gate potential is decreased to -1.2 V, to suppress the edge evanescent incompressible strips. The inset depicts the effect of the gate voltage on the overshoot effect at a fixed excitation current amplitude of 2.5 $\mu$A.}}
\end{figure}

 Next we study a narrow Hall bar of 3 $\mu$m width which is electrostatically defined by metallic surface gates (see inset of Fig.~\ref{fig4}). By tuning the gate voltages it is possible to adjust the carrier density gradients at the edges of the Hall bar, and hence, to manipulate edge and bulk incompressible strips. Fig. \ref{fig4} shows the Hall resistance as a function of the magnetic field for various values of the imposed current. These measurements have been performed at a relatively high temperature of 1.7 K. In Fig.~\ref{fig4}, all gate voltages have been set to $V_g =$ -0.3 V (which is just below the pinch-off value at which the 2DES below the gates is completely depleted). This gate voltage close to the pinch-off value offers the smoothest possible edge confinement (and smallest density gradient) of a working Hall bar. According to  Eq.~\ref{eq:ak_width}, we therefore expect odd filling factor bulk strips to be narrower than even filling factor edge strips (similar as in shallow etched samples, but more pronounced, compare sketch in Fig.~\ref{fig1}a). As for the shallow etched narrow sample (Fig.~\ref{fig2}) we observe an overshoot effect at the low field end of the $\nu=3$ plateau, which points the co-existence of the $\nu=2$ and $\nu=3$ evanescent incompressible strips. Consistent with the high temperature data on the shallow etched sample in the inset of Fig.~\ref{fig2}, the gated sample (at T=1.7 K) also reveals a weakening of the overshoot effect as the current is increased (Fig.~\ref{fig4}). The explanation is the break-down of the Hall effect as already discussed above. Fig. \ref{fig5} shows similar measurements on the same sample as in Fig.~\ref{fig4}, but for a much steeper confinement at the edges due to $V_g =$ -1.2 V. Compared to $V_g =$ -0.3 V a larger current is needed to destroy the overshoot.

Fig. \ref{fig5} depicts the Hall resistance as a function of $B$ for various current values. We observe, as expected, that the overshoot effect tends to disappear while increasing the current. In contrast to the situation in Fig. \ref{fig4} for smoother edges, even at the highest current remainders of the $\nu=3$ plateau still exists. We explain this as follows: The edge profile is steeper at large gate voltages, therefore, the $\nu=2$ evanescent strip is washed out faster than the wider bulk strip of $\nu=3$. To test our explanation due to edge profile together with dissipation, we measured the Hall resistance at the $\nu=3$ plateau interval for $V_G=-0.3$ V and -0.6 V and compared with -1.2 V in the inset of Fig.~\ref{fig5}. We observe that at the smallest gate voltage where the edge is smooth and the $\nu=2$ strip is expected to exist, the overshoot effect is smeared out, due to higher dissipation contributed by the outer most strip.

To summarize this subsection, we utilized a gate defined narrow Hall bar to clarify the contribution of the bulk evanescent incompressible region to the overshoot effect by manipulating the side gate potential. We observed that for all gate voltages, the amplitude of the overshoot effect decreases with increasing current amplitude, as expected. However, if the edge is smooth the overshoot effect disappears even at smaller currents compared to a steep edge (large bulk incompressible region).
\section{CONCLUSION AND DISCUSSION}
\label{sec:con}
To explain the resistance anomalies, namely the overshoot effect, observed in two dimensional electron systems in the integer quantized Hall regime is a long standing challenge. In this article, we present magneto-transport measurements on Hall bars with smooth edges which show a strong overshoot effect and study its dependence on temperature, current, Hall bar width and edge profile, investigating the scattering between the edge-edge and the edge-bulk evanescent incompressible strips. Our results support the screening theory of the quantized Hall effect and its interpretation of the overshoot effect in terms of scattering between edge and bulk evanescent incompressible strips. In more detail we observed flat plateaus starting from low temperature and low current, but the overshoot effect becomes more pronounced as either temperature or current is moderately increased. Too high current, however, causes the breakdown of the QHE and with it the overshoot effect. Once the sample width exceeds the typical length scale of disorder induced long range potential fluctuations, the overshoot effect can be observed not only for odd but also for even filling factors which is related to disorder induced modifications of the bulk strips.

\section*{Acknowledgments}
We would like to thank M. Grayson and A. Wild for fruitful discussions on the overshoot effect. We thank M. Lynass, S.C. Lok and J. Horas for preparing the etched and gated samples, respectively. Istanbul University scientific projects department is acknowledged for supporting us under grant IU-BAP:6970, 13523 and 22662. The scientific council of Turkey (T\"UB\.ITAK) is acknowledged under grant no 112T264. E. M. K. acknowledges T\"UB\.ITAK for financial support under grant 2218. S. L. acknowledges support from the Cluster of Excellence Nanosystems Initiative Munich and the DFG via SFB 631, LU818/7-1 and a Heisenberg Fellowship.

\bibliographystyle{apsrmp}
\bibliography{rmp-sample}


\begin{figure}

\end{figure}

\end{document}